\documentclass [12pt,twoside]{article}           % version LATEX2E

\usepackage{epsfig,times,lscape}
\usepackage[usenames]{color}

    \ifnum\count102=1

    \fi

    \ifnum\count102=2
\fi

\newcommand{\nc}{\newcommand}

     \ifnum\count101=1
\nc{\qI}[1]{\section{{#1}}}
\nc{\qA}[1]{\subsection{{#1}}}
\nc{\qun}[1]{\subsubsection{{#1}}}
\nc{\qa}[1]{\paragraph{{#1}}}

\def\qpar{\vskip 2mm plus 0.2mm minus 0.2mm}
\def\qL{\hfill \break}
     \fi 

      \ifnum\count101=2
 \nc{\qI}[1]{\parindent=0mm \vskip 8mm 
{\centerline{\LARGE \color{red}#1}}\vskip 3mm}
\nc{\qA}[1]{\vskip 2.5mm \noindent 
{{\bf\large\color{blue}  #1}} \vskip 1mm \parindent=0mm}
 \nc{\qun}[1]{\vskip 1mm \noindent {\sl #1 }\quad }

\def\qL{\hfill \break}
\def\qpar{\vskip 2mm plus 0.2mm minus 0.2mm}

      \fi
\def\qth{\vrule height 12pt depth 0pt width 0pt}
\def\qtb{\vrule height 0pt depth 5pt width 0pt}

\nc{\qfoot}[1]{\footnote{{#1}}}

\def\qrr{\color{red} }
\def\qbb{\color{blue} }

      \ifnum\count101=1
\def\qbu{\hfill \par \hskip 6mm $ \bullet $ \hskip 2mm}

\def\qee#1{\hfill \par \hskip 6mm (#1) \hskip 2 mm}

      \fi
      \ifnum\count101=2
\def\qbu{\hfill \par \hskip 4mm $ \bullet $ \hskip 2mm}

\def\qee#1{\hfill \par \hskip 4mm (#1) \hskip 2 mm}

      \fi

\def\qparr{ \vskip 1.0mm plus 0.2mm minus 0.2mm \hangindent=10mm
\hangafter=1}

     \ifnum\count101=1 
 
     \fi
     \ifnum\count101=2

  \def\qcitb#1{\noindent \hbox to 102mm{\hfill \small #1} \vskip 1mm}
      \fi

 \def\qpages#1{\count102=0{\loop\advance\count102 by 1
 \null \vfill\eject \ifnum\count102<#1 \repeat}}

\def\qth{\vrule height 12pt depth 0pt width 0pt}
\def\qtb{\vrule height 0pt depth 5pt width 0pt}

\def\qv{\vskip 0.1mm plus 0.05mm minus 0.05mm}
\def\qhu{\hskip 0.6mm}
\def\qhv{\hskip 3mm}

\def\qhw{\hskip 1.5mm}
\def\qleg#1#2#3{\noindent {\bf \small #1\qhw}{\small #2\qhw}{\it \small #3}\qv }

\begin{document}
\thispagestyle{empty}
% --------------------------------------------------------------------

      % Hauts de pages et numerotation

          % Remarque: sans le \protect --> message d'erreur (ordre fragile)
\markboth{{\sl \hfill  \hfill \protect\phantom{3}}}
        {{\protect\phantom{3}\sl \hfill  \hfill}}

% -------------------------------------------------------------------
\color{yellow} 
\hrule height 20mm depth 10mm width 170mm 
\color{black}
\vskip -2.5cm 
\centerline{\bf \Large How can one explain changes}
\vskip 2mm
\centerline{\bf \Large in}
\vskip 2mm
\centerline{\bf \Large the monthly pattern of suicide?}
\vskip 10mm

\centerline{\large Bertrand M. Roehner$ ^{1} $ }

\vskip 10mm
\large

{\bf Abstract}\quad
The monthly pattern of suicides has remained a puzzle ever since
it was discovered in the second half of the 19th century.
In this paper we intend to ``explain'' not the pattern itself
but rather its changes
across countries and in the course of
time. \qL
First, we show that the fairly common idea according to which
this pattern is decaying in ``modern'' societies is not 
altogether true. 
For instance, around 2000, in well urbanized countries like
South Korea or Spain this pattern was still as strong
as it was in France (and other European countries) 
in the late 19th century.
\qL
The method that we use in order to make some progress 
in our understanding is the
time-honored Cartesian approach of breaking up the problem under
consideration
``into as many parts as might be necessary to solve it''.
More specifically, we try two decompositions of monthly suicides:
(i) according to suicide methods (ii) according to
age-groups. 
\qL
The first decomposition points out the key-role of
hanging and drowning. The second shows the crucial role
of the $ 15-20 $ and $ 65+ $ age-groups. \qL
Then, we present
a number of cases in which 
age-group decomposition provides adequate predictions.
It turns out that
the cases in which the predictions do not work are
newly urbanized countries. The discrepancies may 
be due to a memory
effect which induces a time-lag extending over one or two 
generations.\qL
Finally, in the
light of the new results presented in the present paper,
we re-examine the theory proposed by Emile Durkheim.

\vskip 6mm
\centerline{\it First version: 7 August 2014, comments are welcome}

\vskip 10mm
{\normalsize Key-words: suicide, age, seasonal pattern,
Durkheim, urbanization, South Korea.}
\vskip 10mm

{\normalsize 
1: Institute for Theoretical and High Energy Physics (LPTHE),
University Pierre and Marie Curie, Paris, France. \qL
Email: roehner@lpthe.jussieu.fr
}

\vfill\eject

\qI{Introduction}

Suicide numbers display a seasonal pattern which, more or less,
tends to repeat itself annually. This can be seen just by examining
a monthly time-series. It can be observed
with more accuracy by computing
the auto-covariance function and the spectral density. The later
has a main peak for a 12-month period and a smaller, secondary
peak for a 6-month period. Basically, suicide numbers are lowest
in December and highest in April-May-June. Although this pattern
has been known since the second half of the 19th century
(see the work of Morselli (1879) and Durkheim (1897)
we still have no real understanding of this effect.
\qpar
A paper devoted to this question by Swiss researchers 
(Ajdacic-Gross et al. 2005) starts with the following sentence.
``Seasonality in suicide is one of those topics in epidemiology that we
believe to know much about but understand fairly little in
actuality''. This is a lucid assessment with which most researchers
would probably agree. Why is this so?\qL
To our best knowledge, the only comprehensive framework on
which we can rely is the one proposed by Emile Durkheim.
He showed that there is an inverse relationship between the strength
of inter-social links (and particularly family links)
and  the propensity for suicide. This mechanism explains very
well a number of observed facts (see for instance Roehner (2007, 
Part 3). Does it also explain the seasonal pattern? This point
will be discussed at the end of the paper. 
\qpar
As suicide rates show a minimum in December and a maximum in May 
the first idea which comes to mind is to think that there is
a connection with day-length. Although that connection was already
considered by Durkheim, there are two key empirical tests
that could not be performed at that time due to a lack of data.
\qbu Is the seasonal pattern of suicide reversed in the
Southern hemisphere (e.g. in Australia, Chile or Argentina)?
The answer is yes.
\qbu Is the suicide pattern stronger in northern countries
(e.g. Alaska, the northern provinces of Norway and Sweden) which have
sharper seasonal daylength differences? 
The answer is no. The seasonal effect hardly changes with latitude.
Nowadays
it is almost the same in Alaska and in Florida. Yet, it is possible
that there was a substantial difference in the past.
\qpar

The investigation by  
Ajdacic-Gross et al. (2005) mentioned above 
starts by analyzing the decay of the seasonal effect over
the past 125 years. After presenting solid evidence of this
erosion for Switzerland, the authors
conclude by saying that the erosion
may be related to
the ``transformation of a rural society into a modern one''. 
Durkheim had already observed that the effect is stronger
in rural areas than in big cities.
The slow erosion
seen in Switzerland is also observed in most
industrialized countries. For instance, in France
where monthly suicide data are available since 1836
one observes that
the quinquenial peak-low ratio%
\qfoot{This Peal/Low ratio is defined as 
the highest monthly suicide rate
in 12 months divided by the lowest rate. ``Quinquenial'' refers
to the fact that one considers suicide rates which are
5-year averages.} 
defined by Ajdacic-Gross et al.
decreased from 1.88 to 1.25 during the period 1838-1997,
a trend very similar to the one observed in Switzerland:
0.38/century in France versus 0.52/century in Switzerland.
\qpar

The present author shared the belief that the decay 
should be attributed to urbanization 
until
coming across suicide data for South Korea and Spain.
In these countries, in contrast with the cases mentioned so far,
there is still nowadays a strong seasonal pattern (Fig. 1).

%%%% COMPARAISON FRANCE-COREE-IRLANDE-SUISSE
\begin{figure}[htb]
%\centerline{\psfig{width=14cm,figure=FIG/frcor.eps}}
\centerline{\psfig{width=14cm,figure=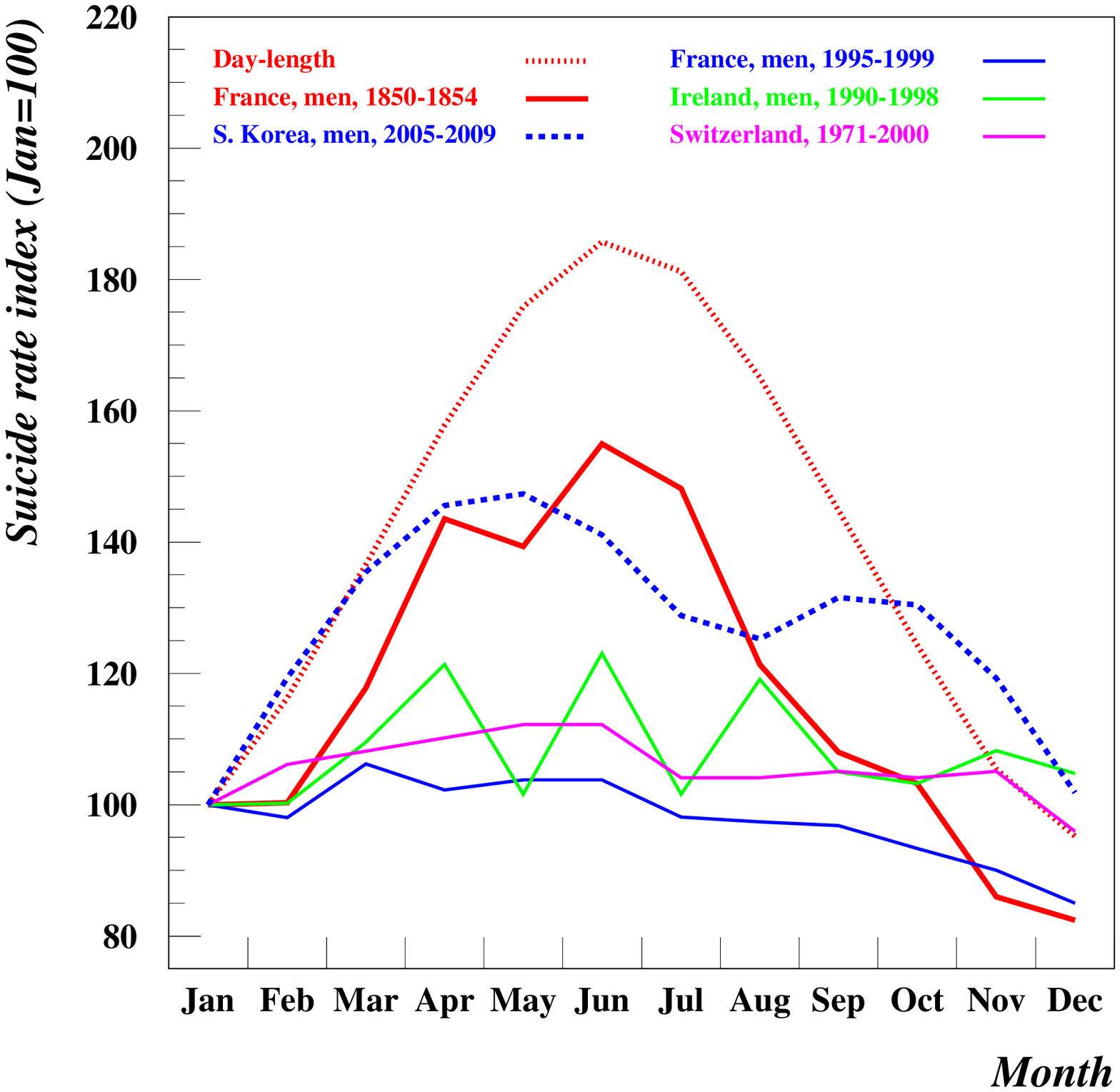}}
\qleg{Fig.\qhu 1 \qhv Monthly suicide pattern in France and Korea.}
{In countries like South Korea there is still a strong seasonal
pattern in spite of a high urbanization rate.
The data for Spain are similar to those for Korea; they
are not shown here for the sake of clarity but are given in Table 1.
All data are for male populations; the data for the female populations
are very similar both in shape and in level.
The day-length data are for a latitude of 50 degrees (city of Aachen
in Germany)}
{Source: France 1850-1854: yearbooks of the
``Compte G\'en\'eral de l'Administration
de la Justice Criminelle''; France 1994-1999: INSERM website;
South Korea 1994-1999: Website of the national statistical 
institute; Spain 1994: website of the national statistical institute.
Ireland: Corcoran et al. (2004).}
\end{figure}
%----------------------------------------------

\qA{How to characterize the seasonal pattern}
The seasonal pattern is completely defined by the 12 monthly
suicide numbers. Of course, it would be useful to be able
to characterize it with a smaller and more transparent
set of parameters. We propose the two following.
\qbu The ratio Peak/Low ($ P/L $) of the highest to the smallest
suicide number
defines the amplitude of the monthly frequencies. 
This metric is fairly standard but it is not sufficient
for it does not tells us whether the monthly changes are
random fluctuations or instead follow a well-defined pattern. 
That is why we add the following indicator.
\qbu The second indicator is the 
correlation ($ cor $) of daylength numbers with monthly suicide
frequencies. It will tell us whether the frequencies display the 
standard pattern with minima in January and December and a
maximum in mid-year. Note that, taken alone, this indicator
would not be sufficient for it does not tell us anything
about the amplitude of the mid-year maximum. Any series that
is more or less symmetrical with respect to its mid-year maximum will
give a correlation close to one, no matter how small the maximum is.
\qpar
If one wishes,
the two indicators can be combined into a single seasonal index defined as:
$ s=[(P/L-1)+cor]Y(cor) $, where $ Y(.) $ is the Heaviside function.
This parameter will be zero as long as the correlation is not positive
and once $ cor $ is positive it will become higher for a profile of
larger amplitude.

%%-----------------------------------------------
\begin{table}[htb]
% LES RESULTATS DES CALCULS SONT SUR LA FEUILLE 1b DU 9 AOUT 2014
% LES CALCULS SONT FAITS DS MAPO#SUICOMPM ET MAPL#SUICIDE2
\centerline{\bf \small Table 1\quad  Magnitude of the seasonal
suicide effect}

\vskip 5mm
\hrule
\vskip 0.5mm
\hrule
\vskip 2mm

\color{black} 
\small

$$ \matrix{
\hbox{Case} \hfill& \hbox{Correlation of} \hfill&
\hbox{Peak/Low}\hfill &\hbox{Seasonal} \hfill&
\hbox{Urbanization}\hfill & \hbox{Ratio of suicides} \hfill \cr
\hbox{} \hfill& \hbox{day-length and} \hfill& &\hbox{index}\hfill &
\hbox{rate}\hfill &\hbox{over 65 to}\hfill  \cr
\qtb
\hbox{} \hfill& \hbox{monthly suicides} \hfill& & & &
\hbox{suicides under 45}\hfill  \cr
\noalign{\hrule}
\qth
\hbox{France 1850-1854} \hfill& 0.94 & 1.82&1.76& 18\% & 0.69 \cr
\hbox{South Korea 2005-2009} \hfill& 0.81 & 1.47& 1.28 & 82\%& 0.70 \cr
\hbox{Spain 1980-2004} \hfill& 0.80& 1.37 & 1.18 &77\% & 0.80 \cr
\hbox{Turkey 2000-2004} \hfill& 0.89& 1.61&1.50& 59\% & 0.20 \cr
\hbox{} \hfill& & & &  \cr
\hbox{France 1995-1999} \hfill& 0.62 & 1.25 &0.87& 78\% & 0.78 \cr
\hbox{Switzerland 1971-2000} \hfill& 0.70& 1.17 & 0.87&73\% & 1.12 \cr
\qtb
\hbox{Ireland 1990-1998} \hfill& 0.47 & 1.23 & 0.70 &62\% &  0.11\cr
\noalign{\hrule}
} $$
\vskip 0.5mm
Notes:
The seasonal index measures the similarity between the
monthly curve and the daylength curve.
The table shows that, contrary to what 
``common sense'' may suggest, 
the magnitude of the daylength effect 
is fairly independent of the urbanization rate.
The numbers in the last column are given in reference
with the coming discussion about the influence of age.
All data are for men, except for 
Switzerland which is for males and
females together.
\qL
{\it Sources: The urbanization rates are from the website
``Trading economics'' except the figure for France 1850-1854
which is from Flora et al. (1987, p. 259). The data for suicides
over 65 and under 45 are from the World Health Organization,
except for Turkey (not available on WHO) which come from the ``Turkish
Statistical Institute''.}
\vskip 2mm
\hrule
\vskip 0.7mm
\hrule
\end{table}
%%-----------------------------------------------

%%-----------------------------------------------
\begin{table}[htb]

\centerline{\bf \small Table 2\quad  Effect on the seasonal pattern
of changes in the frequency of suicide methods.}

\vskip 5mm
\hrule
\vskip 0.5mm
\hrule
\vskip 2mm

\color{black} 
\small

$$ \matrix{
\hbox{} \hfill& \hbox{Peak/Low} \hfill&
\hbox{Correlation}\hfill & \hbox{Weight} \hfill & \hbox{Weight} \hfill \cr
\hbox{} \hfill& \hbox{} \hfill&
\hbox{day-length/}\hfill & \hbox{} \hfill & \hbox{} \hfill \cr
\hbox{} \hfill& \hbox{} \hfill&
\hbox{monthly suicides}\hfill & \hbox{} \hfill &
\hbox{} \hfill \cr
\qtb
\hbox{} \hfill& \hbox{1881-1920} \hfill&
\hbox{1881-1920}\hfill & \hbox{1881-1920} \hfill &
\hbox{1969-2000} \hfill \cr
\noalign{\hrule}
\qth
\hbox{Poisoning} \hfill& 1.08 & 0.43 & 4\% & 16\% \cr
\hbox{\bf Hanging} \hfill& \hbox{\bf 1.64}&\hbox{\bf 0.92}& 42\% & 30\% \cr
\hbox{\bf Drowning}\hfill&\hbox{\bf 2.16} & \hbox{\bf 0.92}& 23\% & 11\% \cr
\hbox{Firearms} \hfill& 1.28 & 0.91  & 22\% & 28\% \cr
\hbox{Cutting} \hfill& 1.53 & 0.72 & 5\% & 2\% \cr
\qtb
\hbox{Jumping} \hfill& 1.45 & 0.72 & 3\% & 12\% \cr
\noalign{\hrule}
\hbox{} \hfill&  &  &  &  \cr
\hbox{Total, 1881-1920} \hfill& &  &  &  \cr
\hbox{}\hfill &
\hbox{P/L}&\hbox{P/L}&\hbox{Cor}&\hbox{Cor}\cr 
\hbox{}\hfill &
\hbox{exp.} &\hbox{obs.}&\hbox{exp.}&\hbox{obs.}\cr
\hbox{}\hfill & 1.59 & 1.61 & 0.980 & 0.976 \cr
\hbox{Total, 1961-2000} \hfill& &  &  &  \cr
\hbox{} \hfill& 
\hbox{P/L} &\hbox{P/L}&\hbox{Cor}&\hbox{Cor}\cr
\hbox{} \hfill&
\hbox{exp.} &\hbox{obs.} &\hbox{exp.}&\hbox{obs.}\cr
\qtb
\hbox{}\hfill & \qbb 1.47 & \qbb 1.17 & \qrr 0.97 &  \qrr 0.70 \cr
\noalign{\hrule}
} $$
\vskip 0.5mm
Notes: ``exp.'' means ``expected'', ``obs.'' means ``observed''.
The expected values are computed by combining the monthly
profiles of the 6 methods as observed in 1881-1920 with the
weight of each method observed firstly in 1881-1920 and
secondly in 1961-2000. The expected changes account
for only a small part (26\% for P/L and 5.1\% for Cor.)
of the observed changes. 
\qL
{\it Sources: Ajdacic-Gross (2005) plus personal calculations.}
\vskip 2mm
\hrule
\vskip 0.7mm
\hrule
\end{table}
%%-----------------------------------------------

\qI{Decomposing the phenomenon of suicide into components}

The overall number of suicides in a given area is an aggregate
variable in which several components are bulked together.
These components 
may not necessarily be ruled by the same mechanism. 
Suicide
by poisoning or by hanging may not follow the same rules.
Suicides of 20-year old persons may differ from suicides of
80-year old persons. Combining heterogeneous factors
will make the description opaque and ambiguous. On the contrary,
in any science, whether physics, chemistry, economics or sociology,
one will get a clearer insight by
studying simple components rather than
mixtures. In the following subsections we will try such an approach
in two different ways. Firstly, we make a distinction
between various suicide methods. Secondly, we make a distinction
between different age groups.

\qI{Joint influence of season and suicide methods}

The authors of the paper by Ajdacic-Gross et al. (2005)
investigate the effect of monthly suicide frequencies for
6 possible methods.
It turns out that for drowning and hanging there is a strong 
daylength-like
seasonal pattern (see table 2). Together these two means represent
a weight of 65\% of all the suicides in 1881-1920. 
\qpar
Now, let us for a moment assume
that in the
time interval 1969-2000 these means have been reduced to
a small percentage of all suicides. 
As the other means exhibit but a weak
seasonal pattern, the reduction in hanging
and drowning  would be able to account for
a substantial erosion of the total suicide pattern (as
indeed observed). An explanation of that kind
would constitute a progress in our understanding. 
Let us briefly explain why.
\qpar

In the previous decomposition there are two different
types of factors: 
\qee{1} The monthly profiles of the various suicide methods.
\qee{2} The weight coefficients which describe the contributions
of each method to the global monthly pattern.
\qpar
The monthly profiles are difficult to measure because
one needs a large number of suicides. How many?
Let us assume that one wishes to distinguish between 10 methods 
and between males and females and that one wants at least 200 suicides
in each month. This would require 
$ 200\times 12\times 10\times 2 =48,000 $ suicides.  
In the United States there were some 30,000 suicides in 2001.
In other words, even in a large country such as the US one
needs to combine several years. This is of course even more
necessary for smaller countries. Thus, the study for Switzerland
required data sets combining some 40 years. 
\qpar
On the contrary, the weight coefficients can be measured 
annually without difficulty.
\qpar
If the global seasonal patterns (in different times and countries)
could be accounted for solely by plugging in the appropriate
weight factors, that would mean that the monthly profiles
for each method are more or less constant both in time and across
countries. Such a finding would make the problem much simpler.
It is in this direction that we wish to go.
\qpar

Can the change in the weight coefficients explain the
erosion in the global seasonal pattern?
In 1969-2000 drowning and hanging still represented
41\% of all suicides. The expected change due to the
reduction from 65\% to 41\% 
(taking also into account the changes in the 4 other means)
is given in Table 2; as can be seen, it is too small
to account for the erosion of the seasonal pattern 
that actually occurred. This means that there has been 
changes in the monthly method profiles.
Indeed, Fig. 3 of Ajdacic-Gross et al
(2005) shows that there was a reduction in spring and summer
suicides for hanging, drowning, firearms, cutting and jumping.
These reductions account for the changes not accounted for by the
weight factors. 
 
\qpar

Nonetheless, for the purpose of explaining changes in
the monthly pattern
the previous decomposition has its usefulness.
\qbu For the sake of simplicity we can restrict our attention
to hanging and drowning. Why? Jumping, poisoning and cutting
represent (in 1881-1920) small fractions of all suicides: 
$ 3\%,\ 4\%,\ 5\% $ respectively which means that, 
whatever their specific shapes%
\qfoot{Their $ s $ indexes are  low; for jumping which has
the highest it is only 1.45.}% 
, 
their contributions will
be small anyway. Firearms suicide represent a more important
fraction, namely 22\%, but its monthly curve remains
within a narrow band of $ \pm 10\% $. This band is three times
more narrow than the one for drowning. In other words,
at least in a first approximation one can forget 4 of the
6 classes. 
\qbu {\bf Longitudinal analysis.} In terms of suicide
methods, each country has its own traditions and one
does not expect them to change dramatically in the course
of time. In the previous example of Switzerland one
of the biggest changes was in the percentage of drowning
which fell from 23\% in 1881-1920 to 11\% in 1969-2000
and to 5.9\% in 2000-2004 (Ajdacic-Gross et al. 2008).
\qbu {\bf Transversal analysis.} Precisely 
because each country has its own ways, there are big transversal
differences in suicide methods. Thus, around 2000, in the United 
States firearms and hanging accounted for 60\% and 20\% of
all male suicides whereas in Germany these percentages were 10\% 
and 55\%.  Based on this difference, one would expect a
stronger monthly pattern in Germany than in the United States.
We will see below (Fig. 4) that this indeed the case although
the difference is small.

\qI{Joint influence of season and age}

A preliminary remark is in order to emphasize that it is
essential to correct monthly suicide numbers by taking into
account the length of each month. At first sight it might
seem that a correction of $ 1/30=3\% $ is less than the
statistical fluctuations of the data and may
therefore be considered as unnecessary.\qL
However, that argument does not hold for two reasons.
\qbu In February the correction is $ (31-28)/30=10\% $.
As month-to-month variations are usually less than 10\%, most
non-corrected series would display a spurious dip in February.
\qbu The seesaw irregularities of uncorrected series will
markedly reduce their correlation with smooth
series such as daylength series.

\qA{Germany}

Suicide statistics which give monthly data for different
age-groups are available for several countries. 
We will mostly use German data because they 
give information for as many as 16 age-groups and cover
several years between 2004 and 2011%
\qfoot{For some reason, the age-group $ >90 $ is
not  given in 2005 and 2008. That is why (when we need all 8 years)
we will limit ourselves to 15 age-groups.}%
.
\qpar
Altogether the database includes some 80,000 suicides.
Fig. 2 shows the monthly profile for 15 age-groups.
In most of the age-groups the fluctuations in different years 
are sufficiently random so as to cancel out. Indeed, in many
cases the average curve remains bounded within a narrow
band limited by the two dotted lines. The cases in which the
average breaks out of the $ \pm 10\% $ strip are: $ 15-20 $
(negative deviation) and $ 65-90 $ (positive deviation).
In other words, the daylength-like pattern is mostly a consequence
of the suicide of elderly people over age 65.
\qpar

 %%%% 15 (=9+6)  GRAPHES PAR CLASSE D'AGE (debut -> 9 graphes)
\begin{figure}[htb]
%\centerline{\psfig{width=16cm,figure=FIG/age1.eps}}
\centerline{\psfig{width=16cm,figure=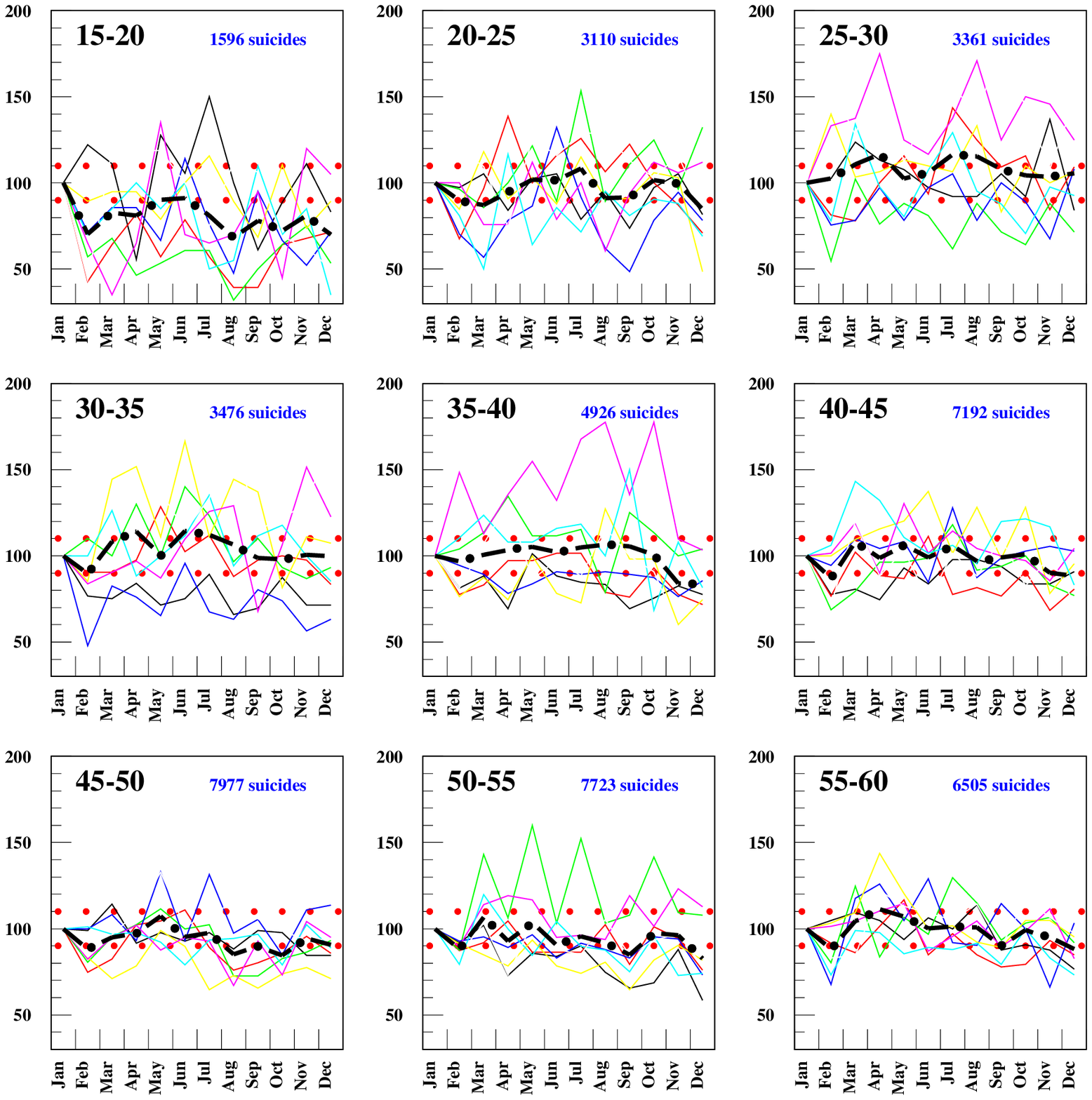}}
\qleg{Fig.\qhu 2a \qhv Suicides by age-group and month.}
{The age-groups include both males and females.
On each graph there are 8 curves (in different colors)
corresponding to the years 2004-2011. The thick dash-dot
curve is the average over those 8 years.}
{Source: The data are from the German Federal Statistical Office,
personal communication from Ms. Silvia Schelo.}
\end{figure}
%----------------------------------------------
%
%%%% 15 (=9+6) GRAPHES PAR CLASSE D'AGE (suite -> 6 graphes)
\begin{figure}[htb]
%\centerline{\psfig{width=16cm,figure=FIG/age2.eps}}
\centerline{\psfig{width=16cm,figure=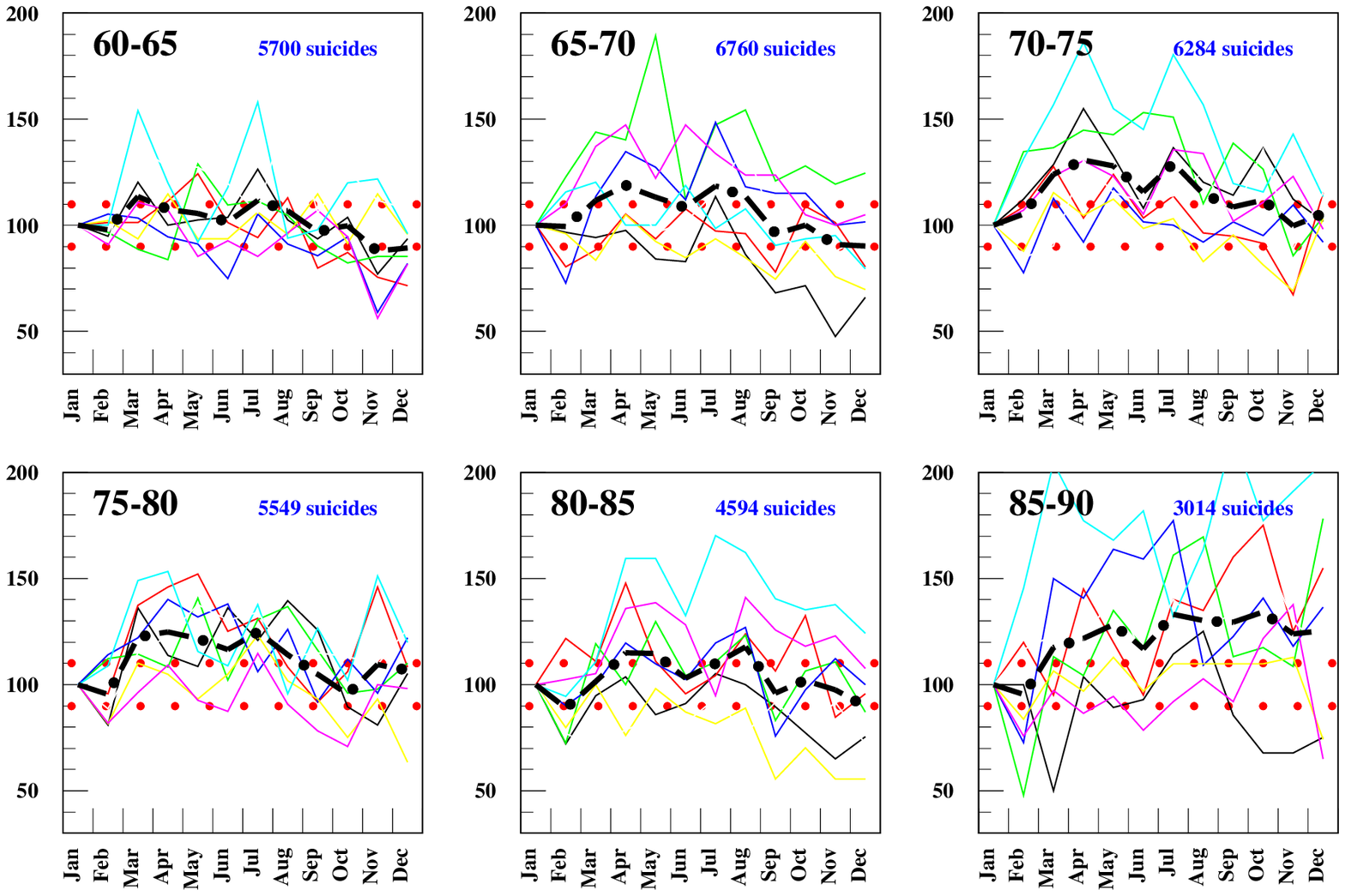}}
\qleg{Fig.\qhu 2b \qhv Suicides by age-group and month.}
{There are only 6 age groups whose monthly profiles
diverge beyond the $ \pm 10\% $ band limited by the dotted
lines, namely $ 15-20 $ and $ 65+ $ which comprises 5 age-groups.
The large fluctuations seen on several of the graphs are due
to small monthly numbers; for instance in the age-group
85-90 the average number of suicides per month is 
$ 3014/(12\times 8)=31 $, had we taken only male suicides
it would be even smaller.}
{Source: The data are from the German Federal Statistical Office,
personal communication from Ms. Silvia Schelo.}
\end{figure}
%----------------------------------------------

\qA{Comparison between Germany and the United States}

On the website of the CDC-NVSS 
(Center for Diseases Control - National Vital Statistics System)
one can download data%
\qfoot{On 15 August 2014 the address was the following:\qL
http://www.cdc.gov/nchs/nvss/mortality/gmwk306.htm}
of suicide numbers per month for 5 different age-groups
(male and female together). These data are available
for 9 years (1999-2007). Altogether they include some
30,000 suicides annually%
\qfoot{This number implies that for a single month
and a single age-group there are on average 
$ 30000/(5\times 12)=500 $ suicides.}
and 270,000 for the 9 years.
Although the dataset is larger than the German dataset,
it is less detailed in terms of age-groups.
\qpar
The interesting fact is that these data lead to results
which are fairly similar to those already described for
Germany (Fig. 3), thus suggesting a validity that extends beyond the
case of one specific country. 
 
%%%% 6 GRAPHES DE COMPARAISON ALLEMAGNE-USA
\begin{figure}[htb]
%\centerline{\psfig{width=16cm,figure=FIG/allusa.eps}}
\centerline{\psfig{width=16cm,figure=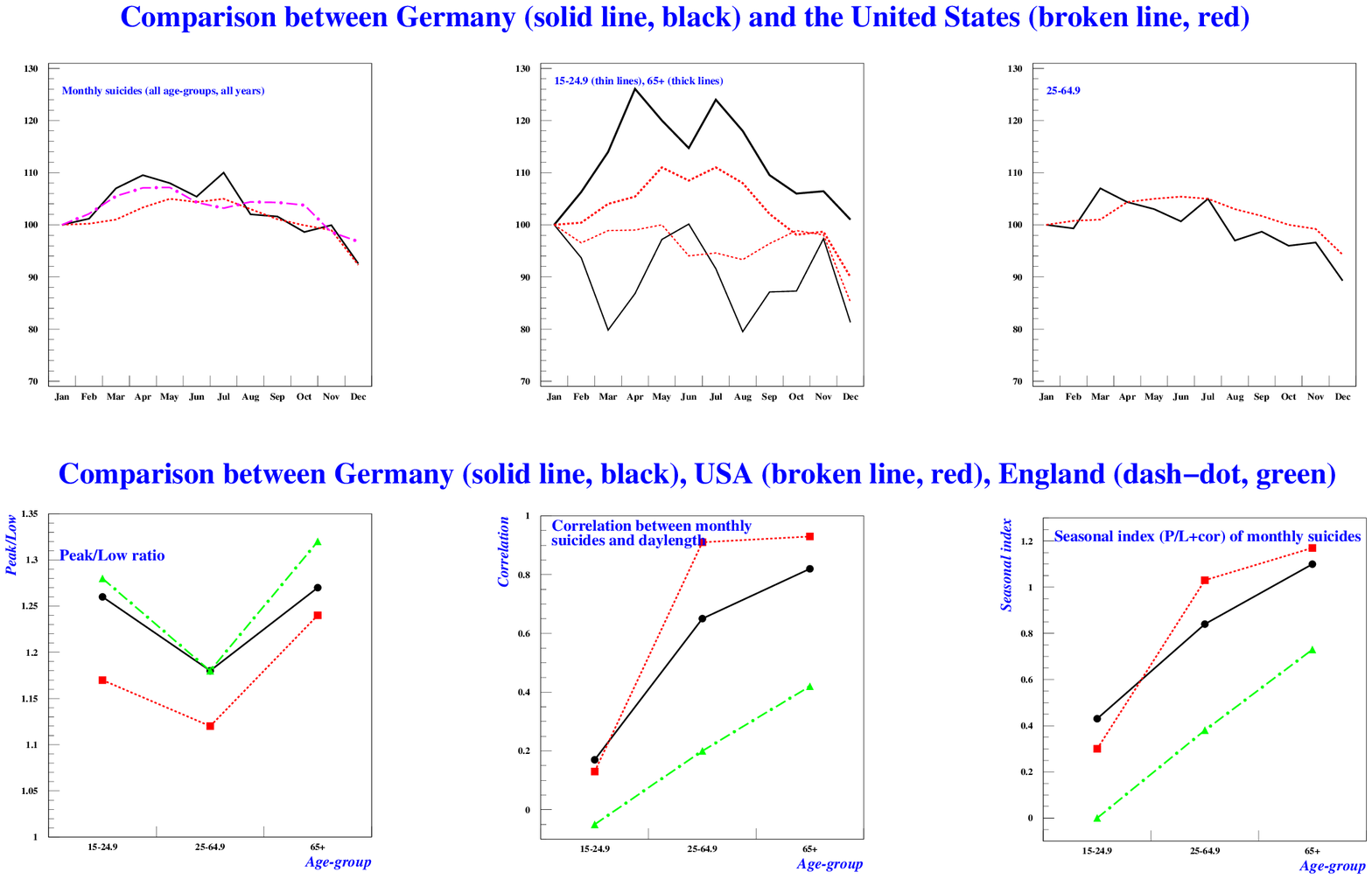}}
\qleg{Fig.\qhu 3 \qhv Comparisons between Germany,
the United States and England.}
{First panel: monthly suicides, all ages. Second panel:
age-group $ (15-24.9) $ (bottom curves), age-group $ (65+) $
(top curves). Third panel: age-group $ (25-64.9) $. 
Fourth panel: $ P/L $. Fifth panel: $ Cor $. Sixth panel: 
Seasonal index $ s $.
Whether in Germany (2004-2011), the US (1999-2007) or 
England (1996-1998),
the young $ (15-24.9) $ and elderly $ (65+) $ age-groups
show monthly profiles which diverge markedly from the
fairly
flat distribution of the middle-age group
$ (25-64.9) $.
In the first panel, for the sake of comparison, we have 
added the curve for the US in 1972-1978 
(dot-dash curve in magenta color). The curves for England were 
not represented in
the graphs of the first line but they are of similar
shape.}
{Sources: The data for Germany are from the Federal Statistical Office.
The US data are from the website of the
National Vital Statistics System (GMWK306) and from 
the ``Vital Statistics of the United States'' annual volumes
(1972-1978) as summarized in MacMahon (1983).
The data for England are from the UK Office
of National Statistics; many thanks to Ms. Anita Brock.}
\end{figure}
%----------------------------------------------
\qpar

\qA{Data for the UK}
Thanks to the help of the British ``Office for National Statistics''
we were able to analyze monthly suicide data by age-group.
The results were in line with those obtained in the cases
of Germany and the United States%
\qfoot{The data are for England, male+female, 1996-1998.}%
.
\qL
Age-group 15-24.9: $ pl=1.28,\ cor=-0.05,\ s=0 $ \qL
Age-group 25-64.9: $ pl=1.18,\ cor=0.20,\ s=38 $ \qL
Age-group 65-84.9: $ pl=1.32,\ cor=0.42, \ s=0.73 $

\qA{Complementary observations}

Naturally, one would like to know whether the regularities
observed for German, US and British data 
also extends to other countries
and other time intervals. In principle such data
should be available in many countries, but they are
not always made available to researchers or sometimes
(as in France) at prohibitive cost.
So far, we could find some complementary information
in the following papers. By and large they confirmed
our previous observations.
\qbu McCleary and collaborators (1991) analyzed some 120,000 individual
suicides which occurred in the United States between 1973 and 1985.
They focused particularly of two age-groups: $ H_1 $: men under 16 and
$ H_2 $: men over 80. In spite of being restricted to very young and very
old persons these age-groups experienced at least 600 suicides 
in any single month which means that their monthly profiles are
reasonably reliable. $ H_1 $ shows a U-shape which goes below the
horizontal line of the uniform distribution.
In other words, it has a negative correlation with the daylength
curve. On the contrary $ H_2 $ has a positive correlation
and a $ P/L $ ratio equal to 1.20. 
\qbu The paper by McCleary et al. (1991) 
was preceded by a paper by Kathleen MacMahon (1982).
As it is not based on individual suicide data it does not
give monthly data at age-group level. It is nevertheless of interest
because it gives the monthly pattern in 1972-1978 that is to say
some 30 years earlier than the 1999-2007 data represented
in Fig. 3. Over this time interval of 27 years
the amplitude of the curve  has {\it not} been reduced:
the peak/low ratio changed from 1.11 to 1.14. Moreover, because
the second curve is more symmetrical with respect to the 
middle of the year, the correlation with the daylength curve
markedly increased from 0.76 to 0.87. As a result, the seasonality
index increased from 0.87 to 1.00. 
\qbu In Hakko et al. (1998) some 21,000 suicides were analyzed
that occurred in Finland during a 16-year study period 1980-1995.
The authors considered only 3 age-groups, namely (i) under 40,
(ii) 40-64 (iii) over 65%
\qfoot{Why did the authors restrict themselves to only 3 age-groups?
The reason is simple. Suppose instead one takes 10 age-groups.
Given that 80\% of the suicides are committed by males
that would mean that on average in each month and each age-group
one would have (i) $ 21000\times (1/10)\times (1/12)\times 0.8=140 $
male suicides (ii) $ 21000\times (1/10)\times (1/12)\times 0.2=35 $
female suicides. Thus one ends with fairly low numbers of events
in each bin.
A rough rule of thumb is that one needs several
hundred suicides in each bin to keep statistical fluctuations
at a ``reasonable'' level (say within $ \pm 20\% $).}%
.
They found that in the age-group $ 85+ $ the monthly suicide had
a $ P/L $ ratio of 1.52.
\qpar

\qA{Reconstruction of monthly profiles}

In order to assess the validity of the simplifying
scheme described above, we will compute the weighted sum $ s(m) $
when it is limited to the selected age-groups.
$$ S(m)=\sum_{k=1}^{16}c_kf_k(m),\quad s(m)=\sum_{k\in A}c_kf_k(m),\quad
S(m)\simeq s(m) $$
$$ \hbox{where: } A=(15-24.9)\cup (65+)=\{1,11,12,13,14,15,16\} $$

Here $ m $ denotes the month, $ k $ is the age-group index, the
$ c_k $ are the weight factors defined as the suicide numbers in
age-group $ k $ divided by the total number of suicides, $ A $ is
the set of selected age-groups, i.e. $ (15-24.9)\cup (65+) $,
$ f_k(m) $ is the monthly profile of age-group $ k $.
\qpar

%
%%%% RECONSTRUCTION
\begin{figure}[htb]
%\centerline{\psfig{width=10cm,figure=FIG/reconstruc.eps}}
\centerline{\psfig{width=10cm,figure=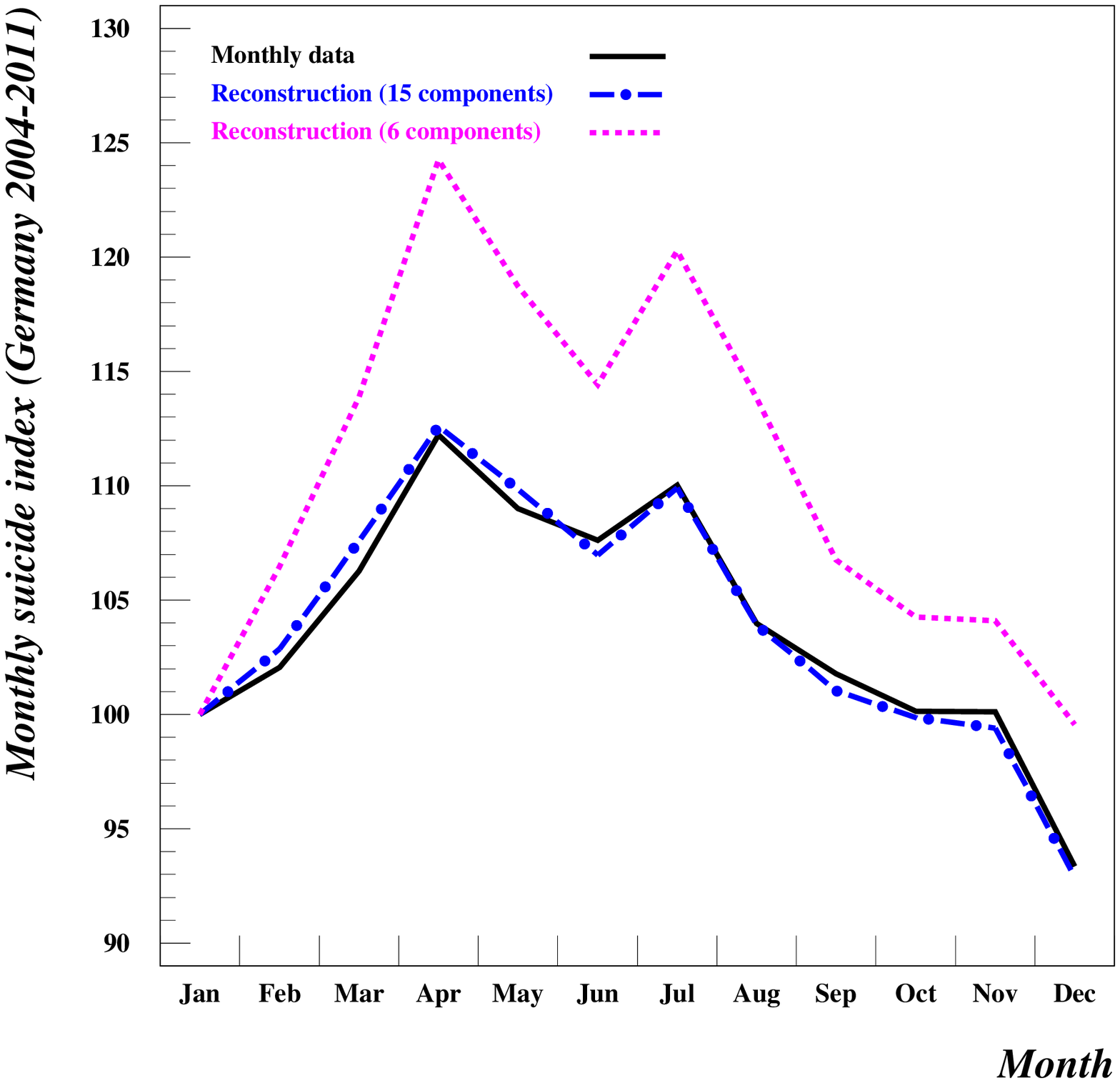}}
\qleg{Fig.\qhu 4 \qhv Testing the reconstruction of the monthly
curve with a reduced number of age-groups.}
{We used only those age groups which contribute significantly
to the monthly pattern, namely $ 15-20 $ and $ 65+ $ (5 age groups).
The reconstructed curve $ s(m) $ (magenta broken line) has the same
shape as the exact curve $ S(m) $ (black solid line); its parameters are:
$ P/L=1.25,\ cor=0.83,\ s=1.08 $ while the parameters of the
exact curve are: $ P/L=1.19,\ cor=0.80,\ s=0.99 $.}
{Source: The data are from the ``German Federal Statistical Office''.}
\end{figure}
%----------------------------------------------

Why is $ S(m) $ not exactly equal to the exact monthly 
curve $ E(m) $? It is because the $ c_k $ are based on the annual
number of suicides; in fact, the monthly ratios have slight fluctuations
with respect to this annual average. 
\qpar

The present test concerned a purely technical of the reconstruction
method. A more fundamental point concerns the stability (in time and
across countries) of the age-group profiles. This point will be tested
in the next section

\qI{Testing transversal predictions}
Let us denote by $ F_1(m), F_2(m) $ the two basic monthly profiles.
If the monthly profiles by age-group (dash-dot curves in Fig. 2ab)
remain fairly stable from one case to another, the predictions
given by the reconstructed monthly curves will be consistent
with observation. If they are not, we can be sure that the 
age-group profiles are not the same in different countries.
This would not be overly surprising. Let us recall that
we have already seen a similar situation in our discussion
of suicide methods.
The prediction
(based on 1881-1920 profiles) for 1969-2000 did not
agree with observation. In this case, as the profiles were known
at each side of the time interval, we could indeed see  
that the monthly profiles of suicide methods
had experienced substantial changes.
\qpar

%-------------------------------------------
% ON SAUTE CE QUI SUIT
\count101=0  \ifnum\count101=1
%%%% RAPPORTS DES NOMBRES DE SUICIDE JEUNES/TOTAL ET JEUNES/VIEUX
\begin{figure}[htb]
\centerline{\psfig{width=10cm,figure=FIG/jeuvieu.eps}}
\qleg{Fig.\qhu xx \qhv Ratios of numbers of suicide by age-group}
{There are two distinct groups of countries.
The group in the lower left-hand-side corner comprises
(from bottom to top): Spain, France, Germany,
South Korea, Switzerland, Hungary ,Japan, China. 
The countries in the upper right-hand-side corner 
are (bottom to top): Finland, USA, UK,
Australia, Russia, Canada, Turkey, Ireland.
Ireland and Finland are obvious outliers; it is not clear why.}
{Source: World Health Organization.}
\end{figure}
%----------------------------------
\fi

\count101=0  \ifnum\count101=1
What should one think of the case of Ireland which is so
obviously an outlier?
Ireland has very low suicide rates for elderly people over 65
but these rates are only low when compared with present-day
rates for young and middle-aged persons; when compared with
Irish suicide rates of 50 years ago these rates are pretty normal.
This suggests the following explanations. Some 50 years ago,
the suicide rates were low because of the influence of the Catholic
church. Nowadays this influence is weaker but for the persons
who were educated and became adults some 50 years ago, that influence
is still prevalent. In other words, while the social environment
has been changing the kind of close social connections developed
by the clergy survived among elderly people.
\qL
One would have more confidence in this argument if it would also
apply to South Korea. Indeed, both countries experienced a rapid
increase in their suicide rate. However, with respect to elderly
persons the situation is quite the opposite for in South Korea
their suicide rate is much higher than the (already high) rates
of young people.
\fi
%%----------------------

\qA{Procedure}
Predictions of monthly patterns were obtained
through the following procedure.
\qee{1} The basic profiles $ F_1(m),\ F_2(m),\ F_3(m) $ 
for the 3 age-groups%
\qfoot{Even though the middle-age component
$ F_2(m) $ is not expected to play a big role, we take it into account
nonetheless.}
$ (15,24.9) $, $ (25,64.9) $, $ (65+) $ were defined by taking the
average of the two cases for which we have good quality data,
namely Germany and the United States.
\qee{2} For each country
3 weight factors $ c_k $ corresponding to the proportions of
the suicides of the 3 age-groups with respect to the total 
number, namely:
$$ c_1=n(15,24.9)/\hbox{All},\ 
c_2=n(25,64.9)/\hbox{All},\ c_3=(65+)/\hbox{All} $$
were computed from WHO statistics.
\qee{3} The 3 components were combined into a single
monthly profile: $ F(m)=\sum_{k=1}^3c_kF_k(m) $ which was then
compared to the monthly profile actually observed. 
\qee{4} In order to estimate the discrepancy between 
predicted and observed profiles we computed their $ P/L $
and $ cor $ indicators. 
For each indicator the differences between
predictions and observations  were plotted (Fig. 5).

\qA{Results}

%%%% PREDICTION-OBSERVATION
\begin{figure}[htb]
%\centerline{\psfig{width=16cm,figure=FIG/predi.eps}}
\centerline{\psfig{width=16cm,figure=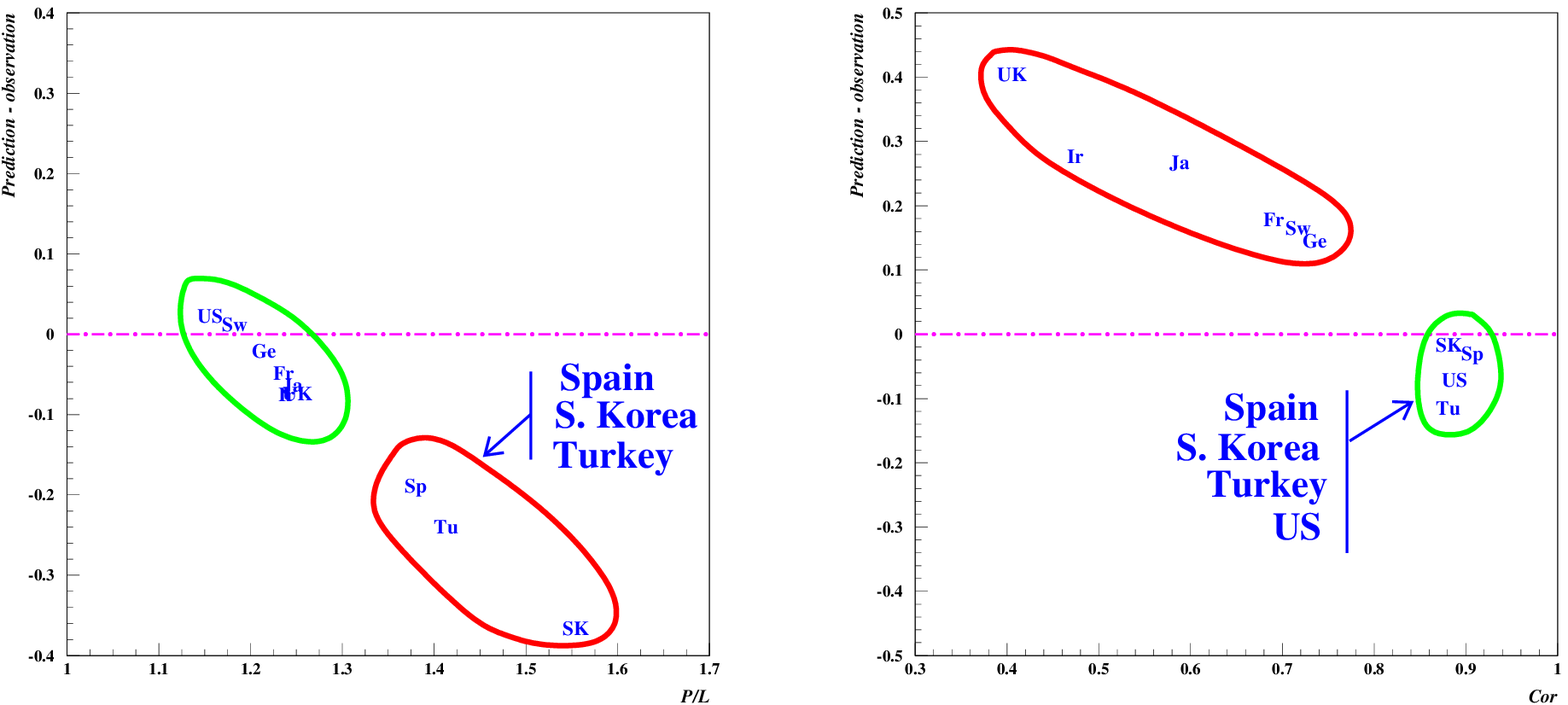}}
\qleg{Fig.\qhu 5 \qhv Discrepancies between predictions
and observations for 10 countries.}
{Left-hand side: $ P/L $, Right-hand side: $ cor $.
Clearly, there are two distinct  clusters.
(i) Spain, South Korea, Turkey 
(ii) France, Germany, Ireland, Japan, Switzerland, UK. 
The United States is in the second subset for $ P/L $ and in the first
for $ cor $.}
{}
\end{figure}
%----------------------------------------------

Spain, South Korea and Turkey are characterized by
high P/L amplitudes. Therefore it was expected that they
could not be accounted for by 
combinations of profiles of fairly low amplitude.
However, at the correlation level the situation
is quite the opposite. That was unexpected.
A tentative explanation may go as follows.
The correlation with daylight is well defined only
when the profile has a clear $ \cap $ or $ \cup $ shape.
When it is fairly flat even the smallest fluctuations
will be able to completely change the correlation.
Thus, the discrepancies seen for low amplitude profile
may just be seen as fairly random fluctuations.

\qA{Spain, South Korea and Turkey}

What makes Spain, South Korea and Turkey so different from
the other cases and in fact similar to what was observed
in European countries one century ago?\qL
This is a key-question.\qL
In order to throw some light it one would need two kinds
of data.
\qbu  Monthly suicide data for one century ago.
They would allows us to see whether or not there was
a downward trend similar to what one observes in
other industrialized countries. 
\qbu Present-day monthly suicide data by age groups.
They would allow us to see whether the amplitude for the
(65+) age group is even higher than for the all-age profile.
\qpar

So far, unfortunately, we were unable to find such data.
In the meanwhile one can tentatively propose the following
explanation.\qL
The explanation relies on two separate effects.
\qbu The first effect is the reduced seasonal effect
in cities. This observation was already made
by Durkheim (1897). He showed
that the seasonal pattern was weaker in big cities than in the
whole country%
\qfoot{He gives evidence for Paris (1888-1892), Berlin (1882-1890),
Frankfurt (1867-1875) and Vienna (1871-1872).
Instead of considering
monthly series he considered three-month averages 
corresponding to the four seasons: Winter (December-February),  
Spring (March-May), Summer (June-August), Fall (September-November). 
In all four cities the corresponding
P/L was lower than in the whole country; the difference was
3\% for Paris, 17\% for Berlin, 20\% for Hamburg and 
23\% for Vienna. The average of these 4 cases is 16\%.}%
.
\qbu The second effect can be referred to as a memory effect.
Observations show that
persons who move from a place $ A $ to a place $ B $ 
will keep the suicide characteristics of $ A $ for a long time.
Thus, people moving from Mexico to the United States will
keep the suicide rate of their country of origin for (at least)
one or two generations. More generally, when immigrants
moved from Europe to the United States their suicide rates
in the United States were closely related
with the suicide rates prevailing in their country of origin
(see Roehner 2007, p. 217-220, the correlation was 0.77). 
Here, we will make the assumption that the same effect
holds not only for the rate but also for the seasonal 
distribution.
\qpar
The consequence of these two effects is that when people
move from the country side (or from small cities) to big cities
they will keep for some time the seasonal characteristics
that they had before arriving in the big cities.
\qpar
One characteristic shared by Spain, South Korea
and Turkey is that 
they experienced rapid economic development and urbanization
during the past decades.
\qpar

Can this explanation be tested in some way?
A place which experienced an economic development similar
to the one in South Korea is the island of Taiwan.
Thus, one would expect a similar seasonal suicide pattern.
Is that the case?\qL
To answer this question one must compare the $ P/L $ and $ cor $
of Taiwan to those of the two subsets in Fig. 4. \qL
For Taiwan (2009-2013) one gets:
$$ \hbox{Taiwan:}\quad P/L=1.35,\quad cor=0.75 $$ 
For Spain, South Korea and Turkey one gets%
\qfoot{The $ \pm $ error bar refers to the confidence interval
for a confidence level of 0.95.}%
:
$$ \hbox{Sp, SK, Tu:}\quad P/L=1.43 \pm 0.09, \quad cor=0.88\pm
0.01 $$
For the 7 other countries, one gets:
$$ \hbox{Ge, US, UK, Fr, Ja, Ir, Sw}\quad P/L=1.20\pm0.02,\quad cor=0.63\pm
0.12 $$

Through its $ P/L $ value Taiwan belongs to the same subset as
Spain, South Korea and Turkey. The conclusion is less clear 
with respect to the $ cor $ variable, but we have already observed
that the correlations in the subset of 7 countries are somewhat random.
\qpar 

Further tests will be possible as soon as reliable monthly
suicide data become available for other
countries which experienced
a rapid increase in their urbanization rate.
\qpar

\qI{What is the mechanism?}

The main objective of the present paper is to focus 
on comparative analysis. That is why so far we did not 
wish to address the question of what is the underlying
mechanism. However, in this concluding section we would
like to discuss some possible mechanisms and particularly
the one suggested by Emile Durkheim (1897).

\qA{Durkheim's conception of the phenomenon of suicide}

In any system composed of individual elements there are two
factors which control the ability of the elements to leave the
system (or to remain inside). 
\qbu {\it Attraction forces} between the elements tend to
prevent the elements from leaving the system.
\qbu At the same time, the elements have their own
independent incentives and impulses which, if strong enough,
can lead them out of the system. For the sake of
brevity we will call this effect the {\it noise factor}.
\qpar
The great achievement of Durkheim was to show that this
model is able to explain several key-properties observed
in suicide statistics. It is important to realize that
the drop-out rate, i.e. the suicide rate, is determined
by the strength of the attraction forces respective to
the strength of the noise factors. 
The most important confirmation
of Durkheim's model is provided by the fact that the
suicide rate becomes higher when the strength of social
links (and particularly family links) decreases. It is
about 3 to 4 times higher for a bachelor than for 
a married man who has several children. 
In this case the effect relies on a change in the
attraction forces whereas the noise factor can be supposed
unchanged.
\qpar
An opposite case is to keep the attraction forces unchanged
and to raise the noise factor. It is on such a mechanism
that Durkheim's explanation of the seasonal pattern relies.

\qA{Durkheim's explanation of the seasonal pattern}

By analyzing the daily distribution of suicides
over one week Durkheim
observed that the number of suicides is 
lowest on Sunday and highest on Monday. This finding was
confirmed by many subsequent studies. For instance, 
MacMahon (1982, p.746) 
found that in the United States between 1972 and 1978, Sunday
had 3\% less suicides than the weekly average whereas Monday
had 8\% more which corresponds to a P/L ratio of 1.11
\qfoot{One may wonder how this rule is changed in countries
in which the day of rest is not Sunday. Modan et al. (1970)
reported that in Israel the number of suicides is highest on
Sunday and lowest on Friday. However, this study is based
on only 322 suicides (corresponding to
1962-1963) which means that for each week-day there were
only $ 322/7=46 $ suicides. That will give high fluctuations
for the daily numbers.}
. 
Based on his observation about daily suicides, Durkheim suggested
the following mechanism.
\qbu He made the assumption that the daily and monthly effects
should be explained by the same mechanism. Although plausible,
this assumption may or may not be true. At this point we do not know%
\qfoot{In support of Durkheim's hypothesis one can observe
that according to our observations for the United States in 1992
both the Sunday-Monday effect  and the
monthly effect are strongest in the age group over 65 and 
almost non-existent in the youngest age-group.
However, in a paper by Ohtsu et al. (2009) one learns
that the Monday-Sunday ratio in Japan (2003) is 1.55
for persons aged 15-64 and 1.49 for the whole population. 
This implies that the ratio for the population over 65
must be substantially smaller than 1.49. Such a 
result is therefore at variance with our observation 
concerning the United States. Further investigations
are required.}% 
.
\qbu Because there is such an obvious contrast
between Sunday and Monday in terms of economic activity,
Durkheim concluded that this is the determining factor 
which explains seasonal and daily variations.
\qpar

In short, Durkheim assumed that the noise factor is in
proportion of economic activity. Assuming stable
attraction forces, higher economic agitation will
of course lead to more suicides. 
This gives a nice explanation
of two of our previous observations.\qL
Because agricultural activity in the fields almost stops during
the winter season that will result in a low suicide rate
in December-January-February as is indeed observed.
In big cities, most economic activities do not
depend upon seasonal weather conditions, but a few do,
as for instance activity in the construction sector.
Thus, the seasonal pattern will be of smaller amplitude
in big cities as seen above.

\qA{Objections to Durkheim's explanation}

At first sight the most obvious objection would be the 
persistence of the seasonal pattern in
highly urbanized countries like South Korea, Spain or Taiwan.
However, if one accepts the time-lag explanation suggested
above, that objection vanishes.
\qpar

In support of his explanation, Durkheim tried
to DEFINE a statistical variable that can
possibly measure ``economic agitation''. For that purpose, 
he used the number of accidents taken as a proxy 
of transportation activity, itself a proxy
of economic activity.  
In the data
that Durkheim presents for Italy (1886-1888) the  maximum 
occurred in Summer and it was
17\% higher than the yearly average. 
Durkheim was satisfied that this Summer maximum
was fairly consistent with the Spring-Summer maximum of 
suicides. 
\qpar

However, if one tries to repeat this test with present-day
death data due to accidents
one finds that there is no correlation between
monthly accidents and monthly suicides. Just to
take one example, in Taiwan (2012-2013), monthly accidents%
\qfoot{Corresponding to the items V01-X59 + Y85-Y86
of the ICD10 classification of causes of death; these 
causes refer mostly to deaths in various forms of transportation,
thus giving an appropriate measure of economic activity.}
have a correlation of -0.17 with monthly suicides.   
In 2012 the maximum of the accident 
series was in March and in 2013 in
July but these maxima were fairly unclear in the sense that the 
series shows fairly random fluctuations.
\qpar

It can of course be argued that the number of accidents is not 
an appropriate measure of ``economic agitation''. This raises the
question of how this agitation should be defined.
\qpar

Should one consider 
the fact that the seasonal effect is stronger for elderly
persons than for middle-age persons as an objection?
Not necessarily. Elderly persons have usually a higher
suicide rate than middle-age persons which means that their
social ties are weaker. Therefore it is not surprising that
they will be more affected by the noise factor.\qL
What is less clear is why the pattern of young people should
be so different. In Durkheim's framework one must assume 
that the higher noise factor in Spring
and Summer keeps them inside the system.
This seems quite surprising and for the moment we have 
no explanation to offer. 
\qpar

At this point we will not try to propose an alternative 
explanation. We think that more comparative work is needed
before an explanation can emerge.

\qI{Conclusions}
First we have shown that taken alone the rate of urbanization
cannot explain the decay of the seasonal
suicide pattern that one observes in 
industrialized countries over the past 150 years.
Spain, South Korea, Taiwan and Turkey do not follow this rule.
In order to throw new light on the question, we
decomposed suicides into more homogeneous components.
Following Ajdacic et al. (2005) we first tried a
decomposition according to methods of suicide.
Secondly, we performed a decomposition into age-groups.
The cases of Germany, the US and the UK (the only countries
for which we were able to get monthly data by age-group)
showed that the age-group decomposition leads to robust
consistent results across countries.
As a confirmation, it was shown that a reconstruction procedure
gives acceptable predictions except for the
four countries already mentioned. 
\qpar
Tentatively, we have proposed a time-lag
memory effect to account for the survival of the seasonal
suicide pattern in newly urbanized societies.
\qpar
Finally, in an attempt to understand the mechanism
responsible of the seasonal pattern, 
we have discussed the explanation proposed by Emile Durkheim.
Coupled with the time-lag memory effect, it appears fairly
satisfactory although further investigation is required
to make clear the precise meaning of
Durkheim's  ``agitation'' variable.

\vskip 10mm
{\bf Acknowledgments}\qL
The author would like to express his gratitude to Ms. Silvia
Schelo of the ``Statistisches Bundesamt'' (Germany's Federal Statistical
Office) for kindly providing data for Germany
and to Ms. Anita Brock of the Mortality Branch of the
UK Office of National Statistics for the data about England.

\vskip 10mm
{\bf References}

\qparr
Ajdacic-Gross (V.), Bopp (M.), Sansossio (R.), Lauber (C.), Gostynski (M.),
Eich (D.), Gutzwiller (F.), R\"ossler (W.) 2005: 
Diversity and change in suicide seasonality over 125 years.
Journal of Epidemiology and Community Health 59,967-972.

\qparr
Ajdacic-Gross (V.),Weiss (M.G.), Ring (M.), Hepp (U.), Bopp (M.), 
Gutzwiller (F.), R\"ossler (W.) 2008: Methods of suicide:
international patterns derived from the WHO database.
Bulletin of the World Health Organization 86,9,726-732.

\qparr
Corcoran (P.), Reilly (M.), Salim (A.), Brennan (A,), 
Keeley (H.S.), Perry (I.J.) 2004: Temporal variation
in Irish suicide rates. 
Suicide and Life-Threatening Behavior 34,4-429-438.

\qparr
Durkheim (E.) 1897: Le suicide. Etude de sociologie. F. Alcan, 
Paris.\qL
[A recent English translation is: ``On Suicide'' (2006),
Penguin Books, London.]

\qparr
Flora (P.), Kraus (F.), Pfenning (W.) 1987: State, economy, and
society in Western Europe 1815-1975. Volume 2. Campus Verlag,
Macmillan Press, St James Press, Frankfort.

\qparr
Hakko (H.), R\"as\"anen (P.), Tiihonen (J.) 2007: 
Seasonal variation in suicide occurrence in Finland.
Acta Psychiatrica Scandinavica 98,2,92–97.

\qparr
MacMahon (K.) 1983: Short-term temporal cycles in the frequency of
suicide.United States 1972-1978. American Journal of Epidemiology
117,744-750.

\qparr
McCleary (R.), Chew (K.S.Y.), Hellsten (J.J.), 
Flynn-Bransford (M.) 1991: Age- and sex-specific cycles in
United States suicides, 1973 to 1985. 
American Journal of Public Health 81,11,1494-1497.

\qparr
Morselli (E.A.) 1879: 
Il suicidio, saggio di statistica morale comparata", Milan, Dumolard.
An English translation was published in 1882 under the title:
Suicide: an essay on comparative moral statistics.
D. Appleton and Company.

\qparr
Ohtsu (T.), Kokase (A.), Osaki (Y.), Kaneita (Y.), Shirasawa (T.),
Ito (T.), Sekii (H.), Kawamoto (T.), Hashimoto (M.), Ohida (T.) 2009:
Blue Monday phenomenon among men: suicide deaths in Japan.
Acta Medica Okayama 63,5,231-236. 

\qparr
Pirkko R\"as\"anena (P.), Hakkoa (H.), Jokelainena (J.), Tiihonenb
(J.) 2002: Seasonal variation in specific methods of suicide: a
national register study of 20,234 Finnish people.
Journal of Affective Disorders 71,51–59.

\qparr
Roehner (B.M.) 2007: Driving forces in physical, biological and
socio-economic phenomena.
A network science investigation of social bonds and interactions.
Cambridge University Press, Cambridge.\qL
[The third part of the book is about suicide.]

\end{document}